# THE ROLE OF CHAOS IN THE CIRCULARIZATION OF TIDAL CAPTURE BINARIES. I. THE CHAOS BOUNDARY

Rosemary A. Mardling

Mathematics Department, Monash University,
Clayton, Victoria, Australia, 3168

*email:* r.mardling@maths.monash.edu.au

ABSTRACT

A self-consistent model for binary evolution devised by Gingold & Monaghan (1980) is used to show that two distinctly different types of behaviour are possible for close eccentric binaries. The model is based on a linear adiabatic normal mode analysis of the problem, which allows detailed examination of the transfer of energy from the orbit to the tides. We show that for most binaries, energy is exchanged quasi-periodically, with the system regulating itself so that the maximum tidal energy always remains small, and no circularization takes place. In contrast, for a range of eccentricities and periastron separations, *chaotic* behaviour prevails, with the eccentricity following a random walk and with the energy transferred to the tides during a single periastron passage being up to an order of magnitude larger than that transferred during the initial periastron passage.

These results have important consequences for the study of tidal capture binaries. The standard model (see, for example, McMillan, McDermott & Taam 1987) assumes that the energy transferred to the tides during a periastron encounter is independent of the oscillatory state of the stars, and that the amount of tidal energy present at any time can be calculated using a formula which (accurately) gives the amount deposited after the first encounter (Press & Teukolsky 1977). The calculations presented here show that a self-consistent treatment is necessary in order to study the dynamical evolution of tidal capture binaries. We conclude that the tidal capture process would not be possible were it not for the existence of chaotic behaviour.

## 1 Introduction

The dynamical history of binaries formed by tidal capture (Fabian, Pringle, & Rees 1975) has been studied only for the first few orbits and for extremely close encounters (for example, Gingold & Monaghan 1980, hereafter GM, Kochanek 1992, Rasio & Shapiro 1991). The energy transferred during the capture process itself was first calculated accurately by Press & Teukolsky (1977) for $n = 3$ polytropes and subsequently by Lee & Ostriker (1986) for $n = 1.5$ polytropes, these providing better models for the low mass main sequence stars found in globular clusters where tidal capture will be most effective. McMillan et al. (1987) considered the energy transfer in more realistic (non-polytropic) models.

Several authors (for example, McMillan et al. 1987 and Ray, Kembhavi, & Antia 1987) have based their studies on the Press & Teukolsky (1977) calculation, and in particular, have extrapolated their results by assuming that the energy transferred during subsequent periastron passages is independent of the pulsational state of the stars. The conclusion is then drawn that the orbit circularizes on a very short timescale ($\tau_{\rm circ} \sim 10$ yr). By then the oscillation energy is of the order of the binding energy of the polytrope, and since the timescale for the most energetic modes to thermalize is much longer than the circularization timescale ($10^4 - 10^6$ yr), the star necessarily responds by expanding, with an estimated order unity increase in the stars radius (McMillan et al. 1987). The conclusion is drawn that most tidal capture binaries result in collision or merger, given that the maximum periastron separation for the process to operate is about 3 stellar radii (for equal mass stars).

The dynamical evolution of tidal capture binaries is clearly important in deciding their fate.

The present work is based on an adiabatic linear normal mode analysis of the problem devised by GM. The self-consistent nature of the model (the total energy is conserved, allowing energy to flow freely between the polytrope and the orbit) allows us to follow the evolution of the system indefinitely (for as long as the energy is conserved to within some tolerance). While numerical methods such as Smoothed Particle Hydrodynamics (SPH - see GM) are good for examining extremely close encounters, particularly those near or at disruption, the present method allows us to examine relatively wide and hence stable orbits for the purposes of studying the formation and evolution of objects such as cataclysmic binaries, low-mass X-ray binaries, pulsar binaries, and black hole binaries, all of which (except, perhaps, black hole binaries) appear plentiful in the cores of globular clusters.

The binary is modelled by a point mass and a polytrope of index $n = 1.5$, representing a compact object and a fully convective low mass Population II star respectively. The mass ratio is arbitrary.

The equations of motion may be derived from a Lagrangian (GM) or directly by considering the stars to be composed of $N$ particles and taking the continuous limit (Mardling 1991). Either way affords a self-consistent treatment (and, of course, yields the same equations), with the stars and orbit free to exchange energy in either direction. A normal mode analysis is then performed in which the velocity and density perturbations are assumed small, and a set of time-dependent ordinary differential equations is derived for the orbit and the oscillation amplitudes of the modes. An arbitrary number of modes may be included.

In Section 2, we review the Lagrangian model devised by GM. In Section 3, we calculate the orbital energy transferred to the tides after the first periastron passage of a parabolic capture orbit for a range of periastron separations and compare the results to those of Press & Teukolsky (1977). Section 4 considers the way the oscillation energy is distributed amongst the modes and the effects of resonance. In Sections 5 and 6, we present two distinct types of orbital behaviour; non-chaotic orbits which are quasi-periodic in eccentricity, and chaotic orbits for which the eccentricity follows a random walk. Section 7 considers for what range of initial periastron separations and eccentricities chaotic behaviour can be expected and plots a boundary between chaotic and non-chaotic orbits. Finally, Section 8 presents a discussion of the results.

## 2 Equations of Motion

The details of the derivation of the Lagrangian for the system may be found in GM. Here we present a summary.

The motion of the system is referred to a non-rotating, non-inertial reference frame with origin at the centre of mass of the polytrope. We assume that the polytrope is adiabatic and has no initial vorticity, and since the forces are conservative, the vorticity remains zero. The velocity of the fluid can therefore be written in terms of a potential, *i.e.* $\mathbf{v} = \nabla\phi$. If $\boldsymbol{\Delta}$ is the position of the point mass relative to the centre of mass of the polytrope and the polytrope and point mass have masses $M_1$ and $M_2$ respectively, the Lagrangian $L$ for the entire system is

$$L = \frac{1}{2}\frac{M_1 M_2}{M_1+M_2}\dot{\boldsymbol{\Delta}}^2 + \frac{GM_1M_2}{\Delta} + GM_2\int_V \rho(\mathbf{r})\left[\frac{1}{|\boldsymbol{\Delta}-\mathbf{r}|} - \frac{1}{\Delta}\right]\mathbf{dr}$$
$$-\int_V \left[\rho\frac{\partial\phi}{\partial t} + \tfrac{1}{2}\rho\nabla\phi\cdot\nabla\phi + \rho U - G\rho\int_{V'}\frac{\rho(\mathbf{r}')}{|\mathbf{r}-\mathbf{r}'|}\mathbf{dr}'\right]\mathbf{dr}. \quad (1)$$

Here $\rho(\mathbf{r})$ is the fluid density at the point $\mathbf{r}$, $\mathbf{dr}$ is a volume element and $U = \kappa n\rho^{1/n}$ is the thermal energy per unit mass, $\kappa$ being a constant and $n$ the polytropic index. The first three terms respectively represent the orbital kinetic energy, the negative of the orbital potential energy and the negative of the interaction energy between the orbit and the fluid, while the remaining terms represent the energy of the fluid. Note that we have not included a term involving $\dot{\boldsymbol{\Delta}}\cdot\mathbf{r}$, which GM later show to represent the linear momentum of the fluid relative to its centre of mass, which of course vanishes.

The Lagrangian may be scaled using the usual Chandrasekhar (1939) scaling:
$\rho = \lambda D$, $\mathbf{r} = \alpha\mathbf{X}$, $\phi = (\alpha^2/\beta)\Phi$, $t = \beta t'$, $M = \lambda\alpha^3 Q$ and $M_2 = sM_1$,
where
$\alpha^2 = (n+1)\kappa\lambda^{(1-n)/n}/4\pi G$, $G\lambda\beta^2 = n/4\pi$, $\lambda$ is the central density so that $D(0) = 1$ and $Q = 4\pi\int_0^{X_0} D(X)X^2 dX$,
$X_0$ being the scaled radius of the polytrope. As in GM, we retain the symbol $\Delta$ for the separation which will now be in units of $\alpha$. We also drop the prime from the scaled time $t'$. Thus $L = (\lambda\alpha^5/\beta)L_0$, where the scaled Lagrangian is

$$L_0 = \frac{1}{2}\frac{s}{1+s}\dot{\boldsymbol{\Delta}}^2 + \frac{nsQ^2}{4\pi}\frac{1}{\Delta} + \frac{nsQ}{4\pi}\int_V D(\mathbf{X})\left[\frac{1}{|\boldsymbol{\Delta}-\mathbf{X}|} - \frac{1}{\Delta}\right]\mathbf{dX}$$
$$-\int_V\left[D\frac{\partial\Phi}{\partial t} + \tfrac{1}{2}D\nabla\Phi\cdot\nabla\Phi + \frac{n^2}{n+1}D^{(1+n)/n} - \frac{n}{4\pi}D\int_{V'}\frac{D(\mathbf{X}')}{|\mathbf{X}-\mathbf{X}'|}\mathbf{dX}'\right]\mathbf{dX}. \quad (2)$$

The density is now written as $D = D_0 + \eta$, where $D_0$ is the Emden density of an unperturbed polytrope and $\eta$ as well as $\Phi$ are assumed to be small.

We now expand $\eta$ and $\Phi$ in terms of the normal modes of vibration of the polytrope:

$$\eta = \sum_{\mathbf{k}} b_{\mathbf{k}}(t)\eta_{kl}(X)Y_{lm}(\theta,\varphi), \quad (3)$$

$$\Phi = \sum_{\mathbf{k}} a_{\mathbf{k}}(t)\phi_{kl}(X)Y_{lm}(\theta,\varphi), \quad (4)$$

where $\mathbf{k} \equiv klm$, $b_{klm}(t) = (-1)^m b^*_{kl-m}(t)$ is the amplitude of the density perturbation, $a_{klm}(t) = (-1)^m a^*_{kl-m}(t)$ is the amplitude of the velocity potential, $\eta_{kl}$ and $\phi_{kl}$ are the radial modes of oscillation as derived in GM (the relationship between these and the standard representation is presented in the Appendix), $Y_{lm}$ are spherical harmonics as defined in Jackson (1975), (with $\int_\Omega Y_{lm}Y_{l'm'}d\Omega = \delta_{ll'}\delta_{mm'}$) $X$, $\theta$ and $\varphi$ are the spherical coordinates of a point in the polytrope and $\sum_{\mathbf{k}} \equiv \sum_{k=1}^{\infty}\sum_{l=2}^{\infty}\sum_{m=-l}^{l}$. It is easily shown that modes with $l = 0, 1$ are not excited. The $k = 1$ modes are the fundamental or $f$-modes, and the $k \geq 2$ modes are the $p$-modes. The $g$-modes are not excited in an $n = 1.5$ polytrope (see, for example, Cox 1980, p238).

The orthogonality condition for the radial modes is

$$\int_0^{X_0} \eta_{kl} \phi_{k'l'} X^2 dX = \omega_{kl} I_{kl} \delta_{kk'} \delta_{ll'}, \tag{5}$$

where $\omega_{kl}$ is the frequency of vibration of mode $\mathbf{k}$. These are accurate to the same number of decimal places as those quoted in Lee & Ostriker (1986). To test the accuracy of the radial normal modes, we follow McMillan et al. (1987) by finding the second derivatives of $\phi_{kl}$ and $\eta_{kl}$ numerically ($\phi_{kl}^{''\,\mathrm{n}}$ and $\eta_{kl}^{''\,\mathrm{n}}$) and comparing them with those obtained directly from the differential equations. This is done by calculating a mass-weighted mean fractional error $\varepsilon_{kl}$:

$$\begin{aligned} \varepsilon_{kl} &\equiv \frac{4\pi \int_0^{X_0} D_0(X) \left(1 - \frac{\phi_{kl}^{''\,\mathrm{n}}}{\phi_{kl}^{''}}\right) X^2 dX}{4\pi \int_0^{X_0} D(X) X^2 dX} \\ &= 1 - \frac{4\pi}{Q} \int_0^{X_0} D_0(X) \frac{\phi_{kl}^{''\,\mathrm{n}}}{\phi_{kl}^{''}} X^2 dX, \end{aligned} \tag{6}$$

which is found to be between $10^{-6}$ and $10^{-2}$ for modes $k = 1, .., 4$ and $l = 2, .., 10$.

Note that since the rotation (toroidal) modes (Unno, Osaki, & Ando 1989, p188) are not excited, $\omega_{kl}$ does not depend on the index $m$ and so is $(2l+1)$-degenerate.

Expanding equation (2) to second order in $\mathbf{\Phi}$ and $\eta$, and using the orthogonality condition equation (5) (as well as that for the $Y_{lm}$), the Lagrangian becomes

$$L_0 = \tfrac{1}{2} \frac{s}{1+s} Q \dot{\mathbf{\Delta}}^2 + \frac{nsQ^2}{4\pi} \frac{1}{\Delta} + nsQ \sum_{\mathbf{k}} \frac{T_{kl}}{2l+1} b_{\mathbf{k}} \frac{Y_{lm}(\pi/2, \psi)}{\Delta^{l+1}} + \tfrac{1}{2} \sum_{\mathbf{k}} (\dot{b}_{\mathbf{k}} \dot{b}_{\mathbf{k}}^* - \omega_{kl}^2 b_{\mathbf{k}} b_{\mathbf{k}}^*) I_{kl}, \tag{7}$$

where

$$T_{kl} = \int_0^{X_0} \eta_{kl} X^{2+l} dX \tag{8}$$

and $I_{kl}$ is defined in equation (5). The equation of motion of the orbit is thus

$$\ddot{\mathbf{\Delta}} = (1+s) \frac{n}{4\pi} \frac{\partial}{\partial \mathbf{\Delta}} \left\{ \frac{Q}{\Delta} + nsQ \sum_{\mathbf{k}} \frac{T_{kl}}{2l+1} b_{\mathbf{k}} \frac{Y_{lm}(\pi/2, \psi)}{\Delta^{l+1}} \right\}, \tag{9}$$

while the amplitudes of oscillation are governed by

$$\ddot{b}_{\mathbf{k}} + \omega_{kl}^2 b_{\mathbf{k}} = \left( \frac{nsQ}{2l+1} \frac{T_{kl}}{I_{kl}} \right) \frac{Y_{lm}^*(\pi/2, \psi)}{\Delta^{l+1}} \tag{10}$$

with

$$a_{\mathbf{k}} = \dot{b}_{\mathbf{k}} / \omega_{kl}. \tag{11}$$

The second term on the right hand side of equation (9) represents the interaction between the tides and the orbit and is responsible for the deviation of the orbit from that of two point masses. The right hand side of equation (10) represents the forcing by the orbit of the oscillations of the polytrope. Equations (9) and (10) are integrated using a fourth-order Runge-Kutta scheme.

### 2.1 Initial Conditions

We start the binary off at apastron ($\psi = -\pi$, $\dot{\Delta} = 0$) or at a separation of about 40 stellar radii for capture orbits and supply the initial eccentricity, $e_0$, and periastron separation, $\Delta_0$, the binary would achieve as two point masses. For the first set of models, we start the polytrope off with zero oscillation energy ($b_{\mathbf{k}} = \dot{b}_{\mathbf{k}} = 0$). We then consider models which start with a prescribed amount of oscillation energy. The initial values for the $b_{\mathbf{k}}$ and $\dot{b}_{\mathbf{k}}$ are obtained from a run which is started with zero initial oscillation energy.

In all models considered, we choose the mass ratio, $s$, to be unity.

### 2.2 Energy and Angular Momentum

The conserved (scaled) energy $E$ and angular momentum $J$ about the centre of mass of the system are shown in GM to be

$$E = \tfrac{1}{2}\frac{s}{1+s}Q\dot{\mathbf{\Delta}}^2 - \frac{nsQ^2}{4\pi}\frac{1}{\Delta} - nsQ\sum_{\mathbf{k}}\frac{T_{kl}}{2l+1}b_{\mathbf{k}}\frac{Y_{lm}(\pi/2,\psi)}{\Delta^{l+1}} + \tfrac{1}{2}\sum_{\mathbf{k}}(\dot{b}_{\mathbf{k}}\dot{b}^*_{\mathbf{k}} + \omega^2_{kl}b_{\mathbf{k}}b^*_{\mathbf{k}})I_{kl}, \tag{12}$$

and

$$J = \frac{sQ}{1+s}\Delta^2\dot{\psi} - \sum_{\mathbf{k}} b_{\mathbf{k}}\dot{b}^*_{\mathbf{k}} imI_{kl} \tag{13}$$

respectively, where $i = \sqrt{-1}$. The first two terms on the right hand side of equation (12) represent the orbital energy, the next term represents the interaction energy between the orbit and the fluid and the last two terms represent the oscillation energy of the fluid. Note that since we take zero initial perturbation to the polytrope, the angular momentum remains in the direction perpendicular to the orbit. The first term of equation (13) represents the orbital angular momentum, while the second term represents the angular momentum transferred to the tides, which remains small relative to the total angular momentum.

The total energy and angular momentum are conserved to at least one part in $10^8$ in all models considered in this paper.

## 3 Comparison with the Press and Teukolsky Model

Press & Teukolsky (1977, hereafter PT) calculated the amount of energy deposited in a star after the first periastron passage of a capture orbit. We first show how their calculation is related to ours, and then compare our model to theirs by repeating their calculation.

In the present units, the rate at which energy is transferred from the orbit to the extended star is given by

$$\frac{dE}{dt} = \int_V D\mathbf{v}\cdot\nabla U\,\mathbf{dX}, \tag{14}$$

where $\mathbf{v}$ is the velocity of a fluid element and $U$ is the gravitational potential of the point mass, given by

$$U(\mathbf{X},t) = \frac{nsQ}{4\pi}\frac{1}{|\mathbf{\Delta} - \mathbf{X}|}. \tag{15}$$

Assuming that the orbit remains parabolic, the amount of energy deposited after one periastron passage is then $\int_{-\infty}^{\infty}(dE/dt)dt$. PT calculate this quantity by writing the velocity in terms of the normal modes of the star, taking the density to be the unperturbed density and assuming that the orbit remains parabolic. In terms of the present model, one may show that in fact,

$$\frac{dE}{dt} = \frac{d}{dt}\sum(\dot{b}_{\mathbf{k}}\dot{b}^*_{\mathbf{k}} + \omega^2_{kl}b_{\mathbf{k}}b^*_{\mathbf{k}})I_{kl}/2. \tag{16}$$

which is the rate at which at which oscillation energy is deposited or removed from the polytrope. If we remove the orbit-oscillation interaction term from equation (9) so that a capture orbit remains parabolic, we find the relative difference between our model and that of PT is 0.006% for a periastron separation of 2 stellar radii, (polytropic index 1.5) increasing to about 0.1% for a periastron separation of 3 stellar radii. On the other hand, if the interaction term is left in, our model predicts about 4% more energy will be deposited after a 2 stellar radii encounter, and this drops to 0.7% at 3 stellar radii.

The latter differences between the predictions of the two models are insignificant for the first periastron passage, but we show in the following section that the PT model can underestimate the energy transfer during a single subsequent passage by up to an order of magnitude if the resulting orbit turns out to be chaotic. On the other hand, we show that for non-chaotic orbits, the energy exchange between the orbit and the polytrope is quasi-periodic, so that energy is not steadily added to the polytrope at each periastron encounter, as was previously thought.

## 4 Mode Energy Distribution and Resonance Effects

Although in this type of study a few $p$-modes are usually included in calculations, they in fact carry little energy. For example, after the first periastron passage of an initially parabolic orbit, the $p_1$ modes ($k=2$) carry less than 0.04% of the total oscillation energy when the periastron separation is $2X_0$, and only 0.0001% when it is $3X_0$. This remains true after subsequent periastron passages.

For the $f$-modes ($k=1$), after the first periastron passage of an initially parabolic orbit with $\Delta_0 = 2X_0$, the energy residing in the $l=3$ and $l=4$ modes is about 13% and 2% respectively of that residing in the $l=2$ mode, while for such an orbit with $\Delta_0 = 3$, the corresponding figures are 5% and 0.5%. This does not necessarily remain true for subsequent orbits. If the orbit happens to be chaotic (see below), it can happen that the $l=3$ mode carries more energy than the $l=2$ mode, and this can become important when a capture orbit ($e_0 = 1$) 'tries' to avoid ionization when in the presence of other stars (see Mardling 1994, hereafter Paper II). On the other hand, if the orbit is not chaotic, the above distribution of energy persists.

Finally, as far as the $m$-modes are concerned, most of the energy in the $l$-modes resides in the $l=m$ modes. After the first periastron passage of an initially parabolic orbit with $\Delta_0 = 2X_0$ ($\Delta_0 = 3X_0$), the energy residing in the $l=m$ modes is 97% (99.5%) of the total oscillation energy, although the distribution is not as severe for less eccentric orbits. As we will show in a future publication (Mardling 1994b), when non-linear mode-mode interactions are taken into account, it is possible for high $l$-modes to be excited, and for these $l$-modes it is the $m=0$ modes that are most excited (compared to other $m$-modes with the same value for $l$).

The tides of stars in close binary orbits are principally raised by the coupling of the orbit and the $l=m$ modes. Consider the term which forces the oscillations in equation (10). Recall that $Y_{lm}(\pi/2,\psi) = c_{lm}P_l^m(0)e^{im\psi}$ (Jackson 1975), where $c_{lm}$ is a constant and $P_l^m$ is a Legendre function of degree $l$ and order $m$. Clearly, resonance effects occur if it happens that $m\dot{\psi} \simeq \omega_{kl}$. This is in fact true for $l=m$ when an orbit is at periastron, as can be seen in Figure 1 which plots curves in $(\Delta_0, e_0)$ space along which $l\dot{\psi}_p/\omega_{kl} = r$ for various $l$, and for $r = 1$ and $r = \frac{1}{2}$ ($\dot{\psi}_p$ may be written in terms of $\Delta_0$ and $e_0$ by using the equation for an unperturbed orbit). Here and elsewhere, the subscript $p$ refers to the value of the orbital variable at periastron.

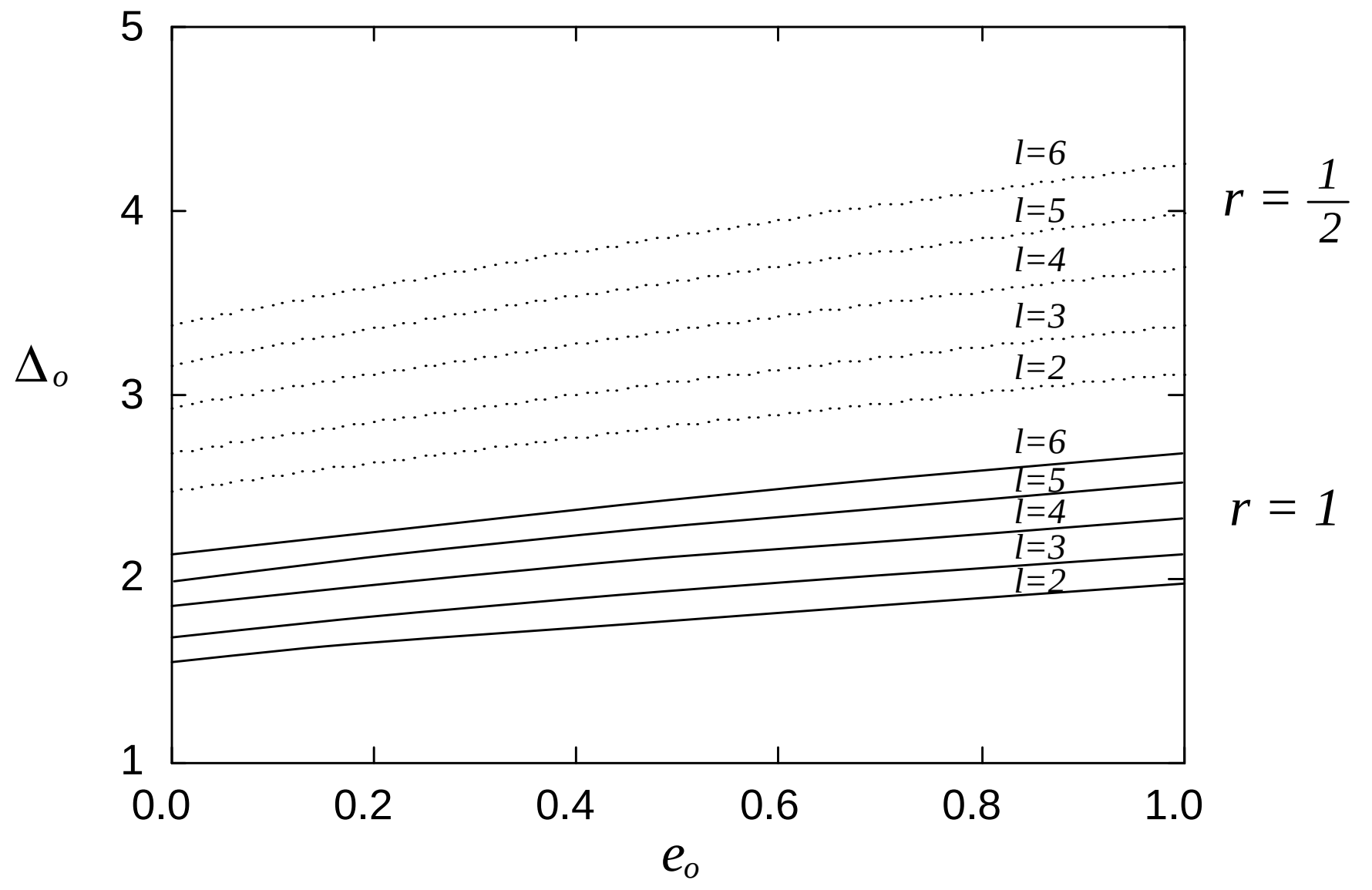

Figure 1: Resonance curves in the $(\Delta_0, e_0)$ plane, along which $l\dot{\psi}/\omega_{kl} = r$ for $r = 1$ and $r = 1/2$ and $l = 2, 3, \ldots, 6$. Here and in the figures which follow, $\Delta_0$ is measured in units of the scaled stellar radius, $X_0$.

## 5 Non-Chaotic Orbits

In order to examine the dynamical behaviour of these systems, we plot the orbital eccentricity against the number of periastron passages. The eccentricity actually varies continuously, with maximum changes occuring near periastron. Nonetheless, the approximate eccentricity can be obtained via $e = (1 - r)/(1 + r)$, where $r$ is the ratio of the apastron separation to the following periastron separation. Figure 2 shows the evolution of the orbit with initial conditions ($\Delta_0 = 3.2X_0, e_0 = 0.7$). The $l = 2$ modes are included, *and the solution is not very sensitive to the inclusion of more modes*: the deviation from this model by one which includes modes up to $l = 4$ is barely discernable after 20 orbits.

Energy is periodically extracted from the orbit and then replaced. The direction of flow of energy depends on the relative phase of the orbit and the oscillations at periastron, with the long term behaviour exhibiting a beating effect. This is in contrast to the assumption made by McMillan et al. (1987) and others that energy is extracted from the orbit at each periastron encounter so that the orbit steadily circularizes.

Figure 3 shows the advance of the apsidal line for this model, which is 26% greater than the classical prediction (see, for example, Schwarzschild 1965). This difference reduces to zero as the separation increases, as indicated in Figure 4 for an eccentricity of 0.4 (all orbits calculated for this figure are non-chaotic).

The example in Figure 2 is representative of the general behaviour of most non-chaotic orbits. Other systems, such as that shown in Figure 5 for the case ($\Delta_0 = 2.87X_0$, $e_0 = 0.2$) exhibit a longer period and a larger amplitude, indicating that more energy is extracted from the orbit than in the typical case, but the important features to keep in mind are that the average eccentricity is constant and that the amplitude of the variation is of the order of the change in eccentricity after one periastron passage for most cases, so that *the tidal energy always remains small* and *no circularization takes place* (at least in the absence of dissipation). These last two points have important consequences for the evolution of tidal capture orbits. This aspect will be examined in Paper II.

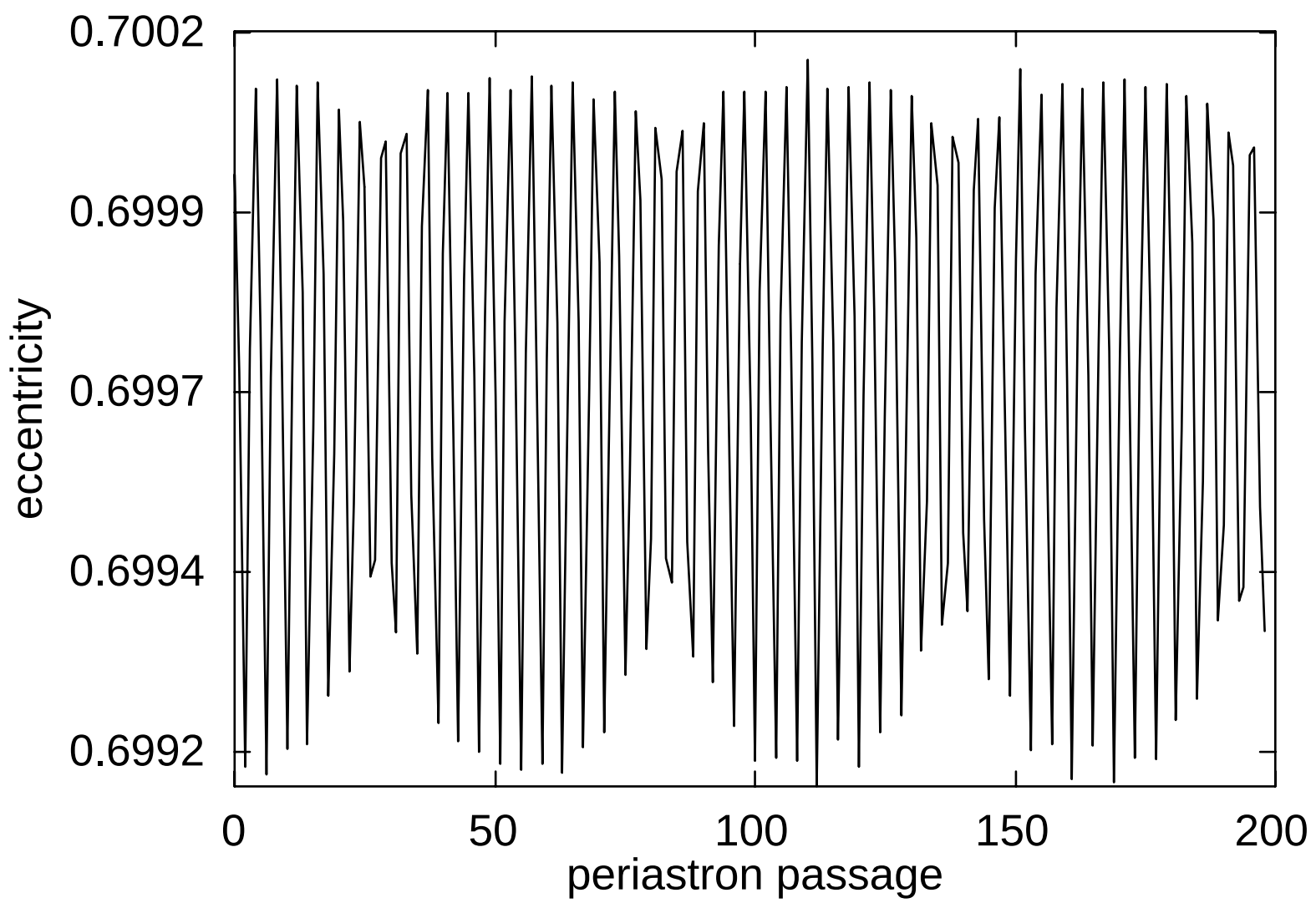

Figure 2: A non-chaotic orbit for which the eccentricity versus periastron passage exhibits beating. Note that the total change in eccentricity is small. Here, $\Delta_0 = 3.2X_0$ and $e_0 = 0.7$.

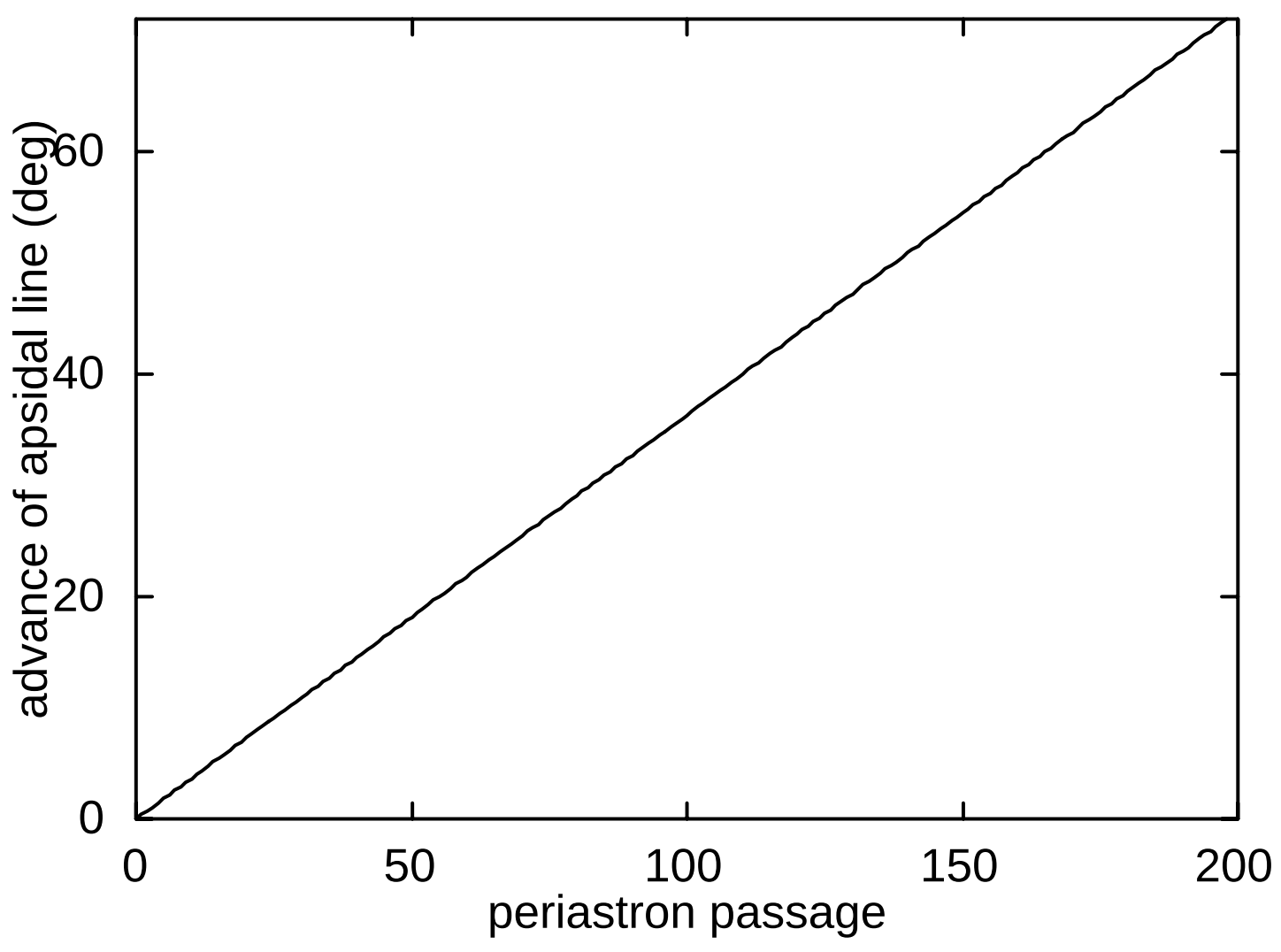

Figure 3: The advance of the apsidal line for the model shown in Figure 2. The rate of advance is constant for non-chaotic orbits.

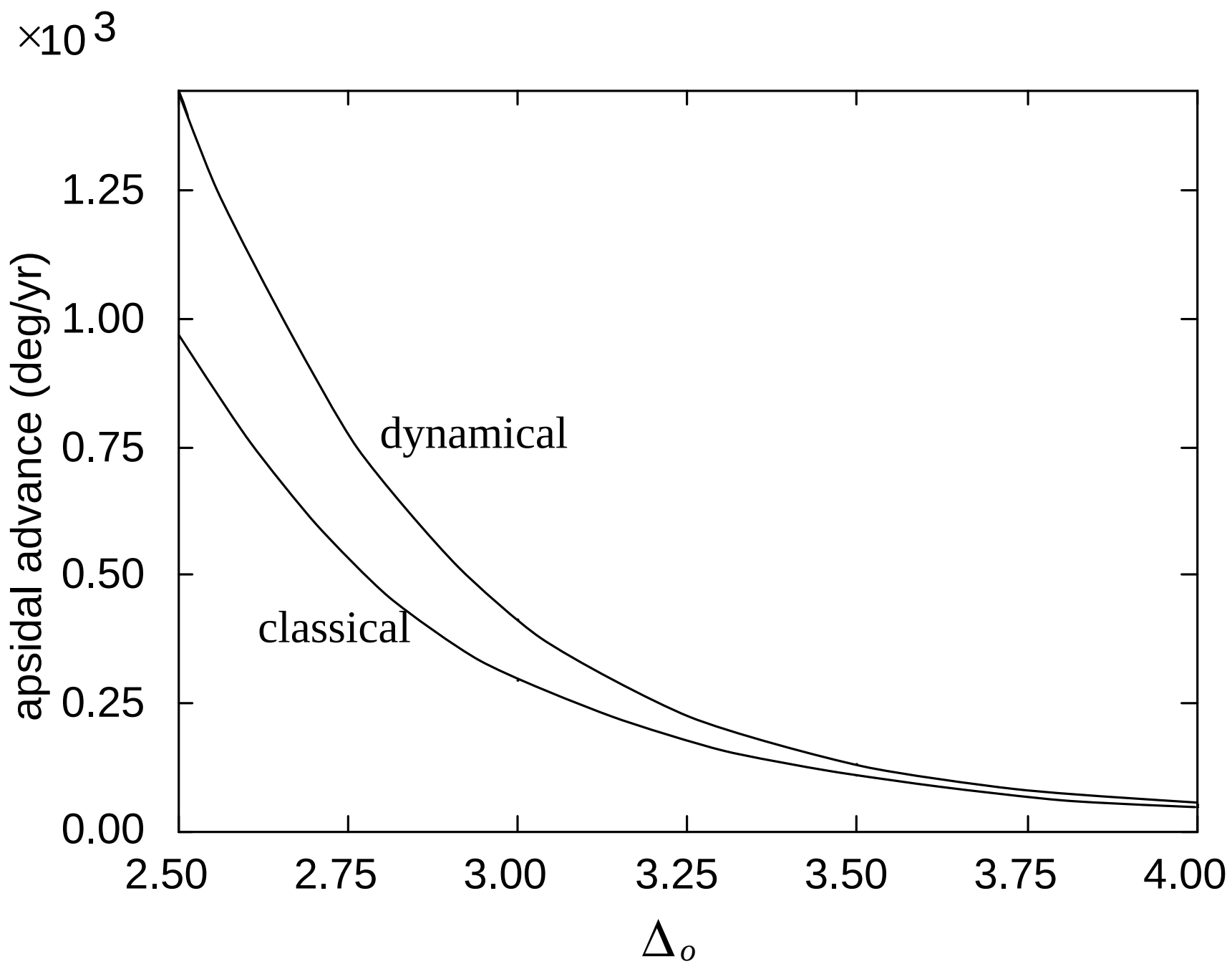

Figure 4: Comparison of the classical and dynamical calculations of the apsidal advance as a function of initial periastron separation, $\Delta_0$ (for non-chaotic orbits only). The fractional difference decreases with increasing $\Delta_0$. Here, $e = 0.4$.

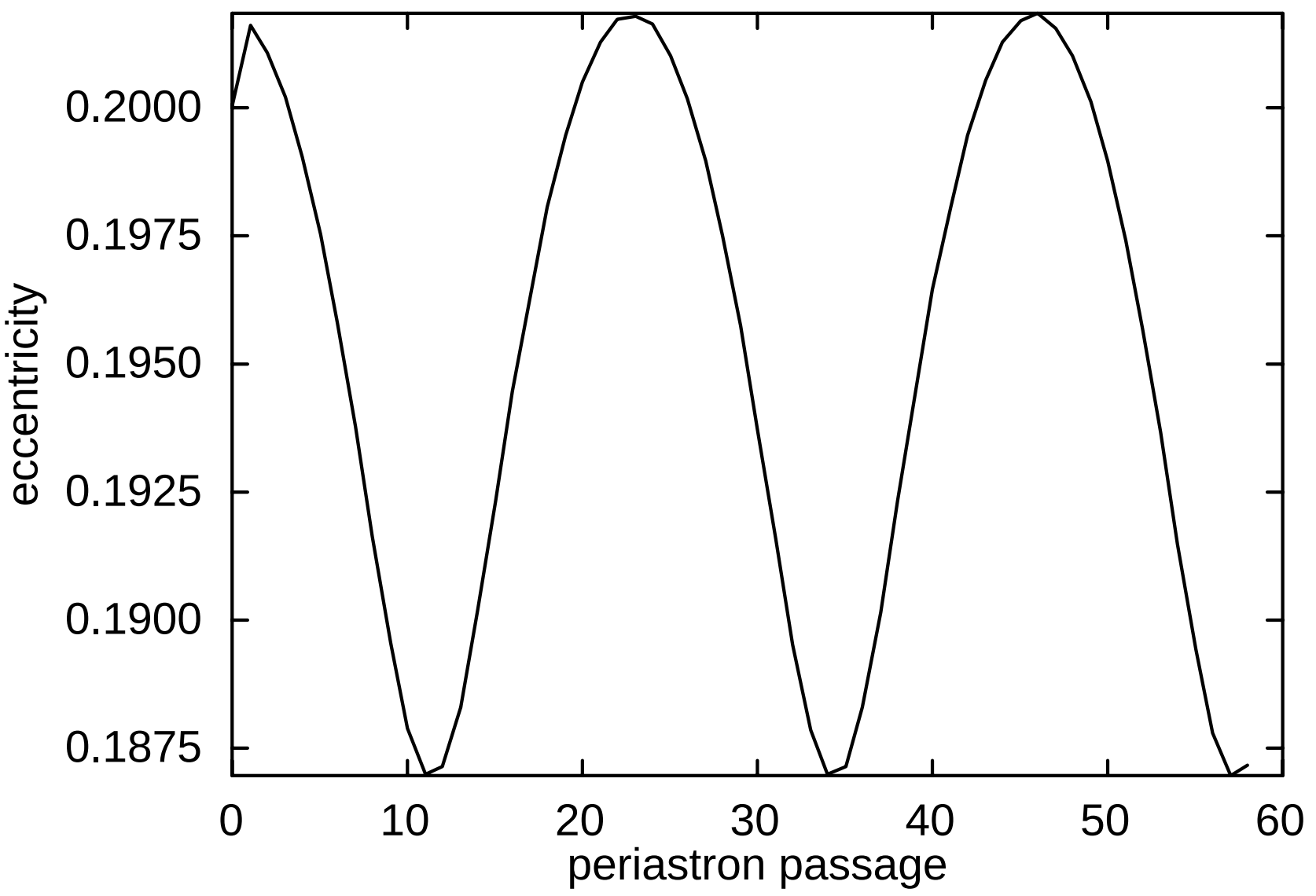

Figure 5: A non-chaotic binary with a long period in the eccentricity. $\Delta_0 = 2.87 X_0$ and $e_0 = 0.2$.

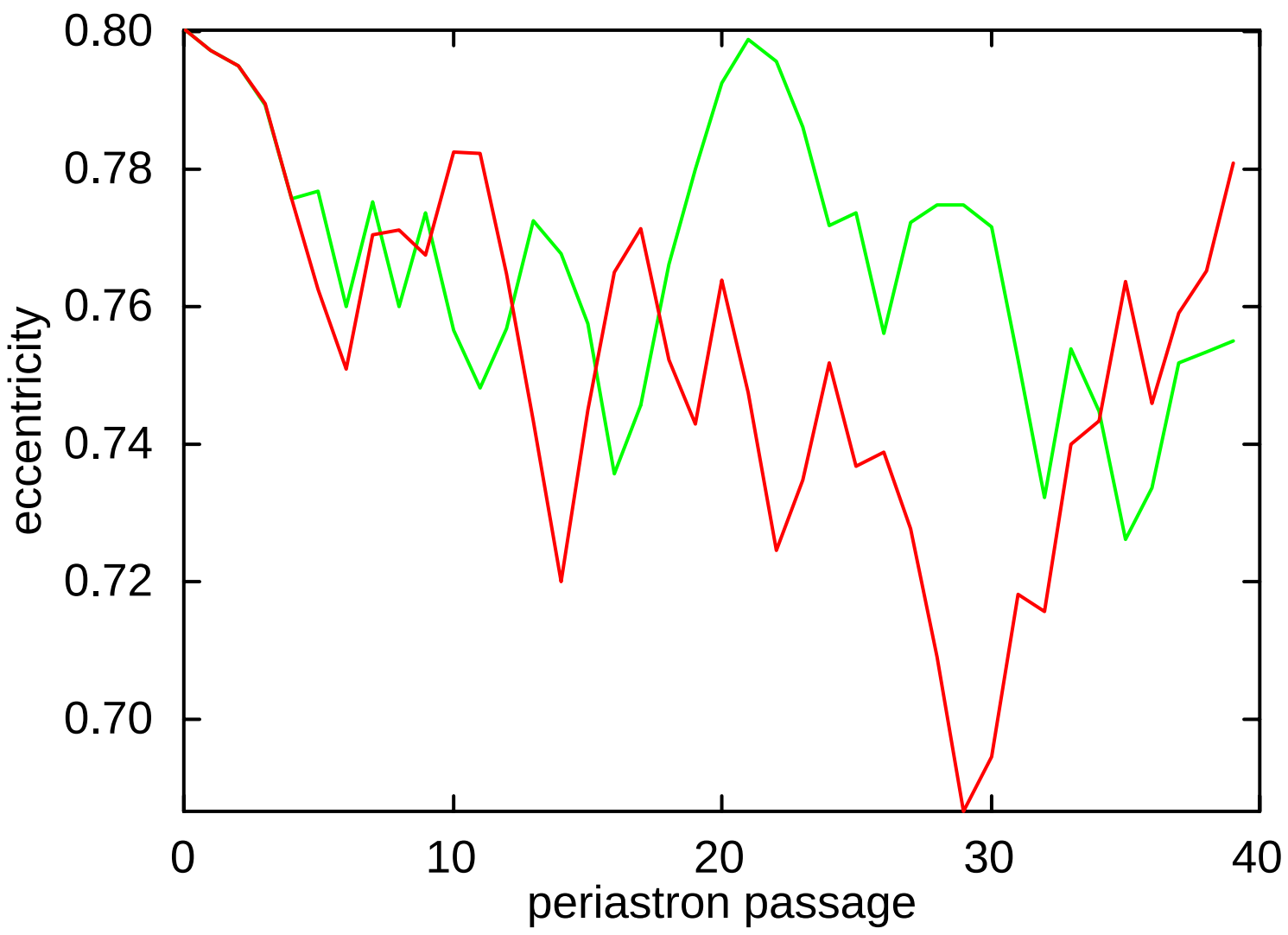

Figure 6: Two initially close chaotic orbits for which ($\Delta_0 = 2.8X_0$, $e_0 = 0.8$) and ($\Delta_0 = 2.80001X_0$, $e_0 = 0.8$). The orbits diverge after the $4^{\rm th}$ orbit. The total change in eccentricity is not bounded as it is for a non-chaotic orbit (but see Paper II).

In Section 7 we will consider for which parameter values $\Delta_0$ and $e_0$ this type of behaviour is found, when we plot the boundary between chaotic orbits and non-chaotic orbits.

## 6 Chaotic Orbits

Figure 6 compares the evolution of the two orbits with initial conditions ($\Delta_0 = 2.8X_0$, $e_0 = 0.8$) and ($\Delta_0 = 2.80001X_0$, $e_0 = 0.8$), while Figure 7 compares the apsidal advance. The behaviour is in stark contrast to that of non-chaotic orbits and clearly depends sensitively on the initial conditions. It is also extremely sensitive to other changes such as the inclusion of more modes, and changes in the timestep used to integrate the equations (and even the machine one performs the calculations on!)

The eccentricity varies over a wide range of values, and any one change in eccentricity after a periastron encounter can be up to an order of magnitude more than the initial change.

Other ways of establishing whether or not a solution is chaotic include determining the Lyapunov exponents of the system (see, for example, Rasband 1990), and plotting a

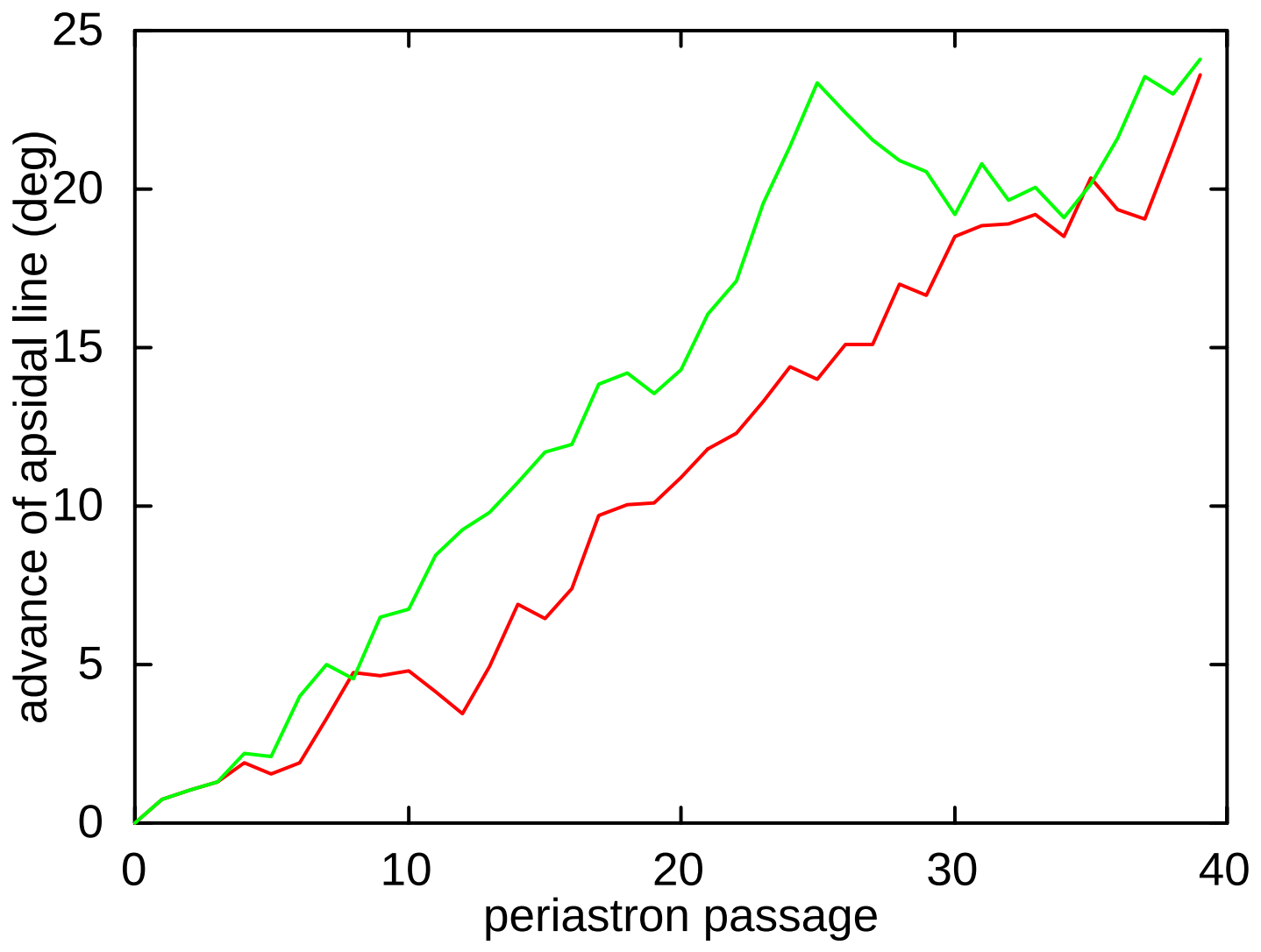

Figure 7: The apsidal advance for the model shown in Figure 5. The rate of advance is not constant for chaotic orbits, and can even be negative.

Poincaré surface of section (Rasband 1990). The number of degrees of freedom is large for this system, even when only the $l = 2$ mode is included. The author has developed a method for calculating the Lyapunov exponents of large systems efficiently (Mardling 1991, 1994c), based on a method devised by Shimada & Nagashima (1979), and independently by Benettin et al. (1980). Applied to the present system with just the $l = 2$ modes included (see Mardling 1991), we find the largest Lyapunov exponent for a typical chaotic orbit to be about 0.01.

## 6.1 Surface of Section

A traditional way of visualizing a system's transition to chaos is to plot its Poincaré surface of section. Normally this is only possible when the system has 3 degrees of freedom, which occurs for 3 dimensional systems and higher dimensional systems for which the number of integrals of the system reduces the degrees of freedom to 3.

If we include only the $l = 2$ $f$-modes, it is possible to plot a surface of section for the present system, even though it has 6 degrees of freedom (after the energy and angular momentum integrals are taken into account).

To see how we may do this, let us first recall how a Poincaré surface of section is drawn for a system with three degrees of freedom. For example, for the four dynamical variables, $x$, $y$, $\dot{x}$ and $\dot{y}$, with integral $f(x, y, \dot{x}, \dot{y}) = E$, where $E$ is a constant, we might plot $x$ versus $\dot{x}$ every time the solution curve crosses the $y = 0$ plane, with $\dot{y}$ being given by the integral. We would then observe whether successive points tended to lie on curves, indicating the existence of a further integral, or even whether the points were mapped to a finite number of points, indicating that the system was completely integrable and that the solution was periodic. Chaotic motion would be indicated if the iterates tended to be space filling.

This method does not normally work for more than three degrees of freedom because in order to examine the behaviour of the orbit as it crosses a two dimensional subspace, we require, for example, several variables to pass through zero coincidentally which is unlikely to occur in most situations. As it happens, a judicious change of variables makes this possible for the present model when we only include the $l = 2$ $f$-mode.

We imagine that our set of dynamical variables is replaced as follows[1]:

$$\{\psi, \Delta, \dot{\psi}, \dot{\Delta}, (b_{\mathbf{k}}, b^*_{\mathbf{k}}), (\dot{b}_{\mathbf{k}}, \dot{b}^*_{\mathbf{k}})\} \Rightarrow \{E, J, e, \dot{\Delta}, \mathcal{I}_{\mathbf{k}}, E_{\mathbf{k}}\}, \tag{17}$$

where $E$ is the total energy given by equation (12), $J$ is the total angular momentum given by (13), $e$ is the orbital eccentricity, $\mathcal{I}_{\mathbf{k}}$ is the interaction energy between the orbit and the $\mathbf{k}^{\mathrm{th}}$ mode and $E_{\mathbf{k}}$ is the oscillation energy in the $\mathbf{k}^{\mathrm{th}}$ mode, with $\mathbf{k} = (1, 2, 0)$, (1,2,-2) and (1,2,2). $\mathcal{I}_{\mathbf{k}}$ and $E_{\mathbf{k}}$ are defined as follows:

$$\mathcal{I}_{\mathbf{k}} = -nsQ\frac{T_{kl}}{2l+1}\frac{1}{\Delta^{l+1}}(b_{\mathbf{k}}Y_{lm} + b^*_{\mathbf{k}}Y^*_{lm}), \tag{18}$$

$$E_{\mathbf{k}} = (\dot{b}_{\mathbf{k}}\dot{b}^*_{\mathbf{k}} + \omega^2_{kl}b_{\mathbf{k}}b^*_{\mathbf{k}})I_{kl}/2. \tag{19}$$

At apastron, $\dot{\Delta} = 0$, and for sufficiently high eccentricity, $\mathcal{I}_{\mathbf{k}} \simeq 0$ so that the orbit and oscillations are essentially decoupled. Thus $e$ and the $E_{\mathbf{k}}$ are approximately constant. We also find that for non-chaotic orbits, the energy in the $m = 0$ modes drops to zero at apastron, so that since $E$ and $J$ are constant, we are left to plot the energy in the (1,2,2) mode against the eccentricity.

Figure 8 shows a series of surface of section plots for models with $e_0 = 0.6$ and $\Delta_0 = 2.845X_0$ to $\Delta_0 = 2.848X_0$. Figure 9 shows the final plot in Figure 8 for later times, while Figure 10 plots the eccentricity against periastron passage for this model, showing how the orbit passes through two non-chaotic phases, seen clearly in Figure 9.

# 7 The Chaos Boundary

We can do a systematic search for chaotic systems in the $(\Delta_0, e_0)$ parameter space by comparing initially very close orbits and seeing if they diverge after the first few orbits. In this way, we can dertermine the 'chaos boundary' above which chaotic behaviour ceases. Figure 11 shows the results of this search. The sections of the curves corresponding to small eccentricities are not of much practical value: models starting with these values for $\Delta_0$ and $e_0$ disrupt when non-linear terms are included in the equations of motion. Intermittant chaotic behaviour is observed above the curves shown: there must exist a curve beyond which all chaotic behaviour has ceased. Nonetheless, the curve shown has practical value, as we show in Paper II, where we consider capture orbits.

All models calculated so far have been started with zero initial oscillation energy. We can also plot chaos boundaries for models with specific initial oscillation energies. These also prove to be useful for examining capture orbits. These are shown in Figure 11 as the terminating curves: they terminate at the point in $(\Delta_0, e_0)$ space which corresponds to zero *total* energy. Thus, of course, the 'zero oscillation energy' chaos boundary terminates at a point on the line $e = 1$.

[1]Recall that $b_{\mathbf{k}}$ and $b^*_{\mathbf{k}}$ are not independent.

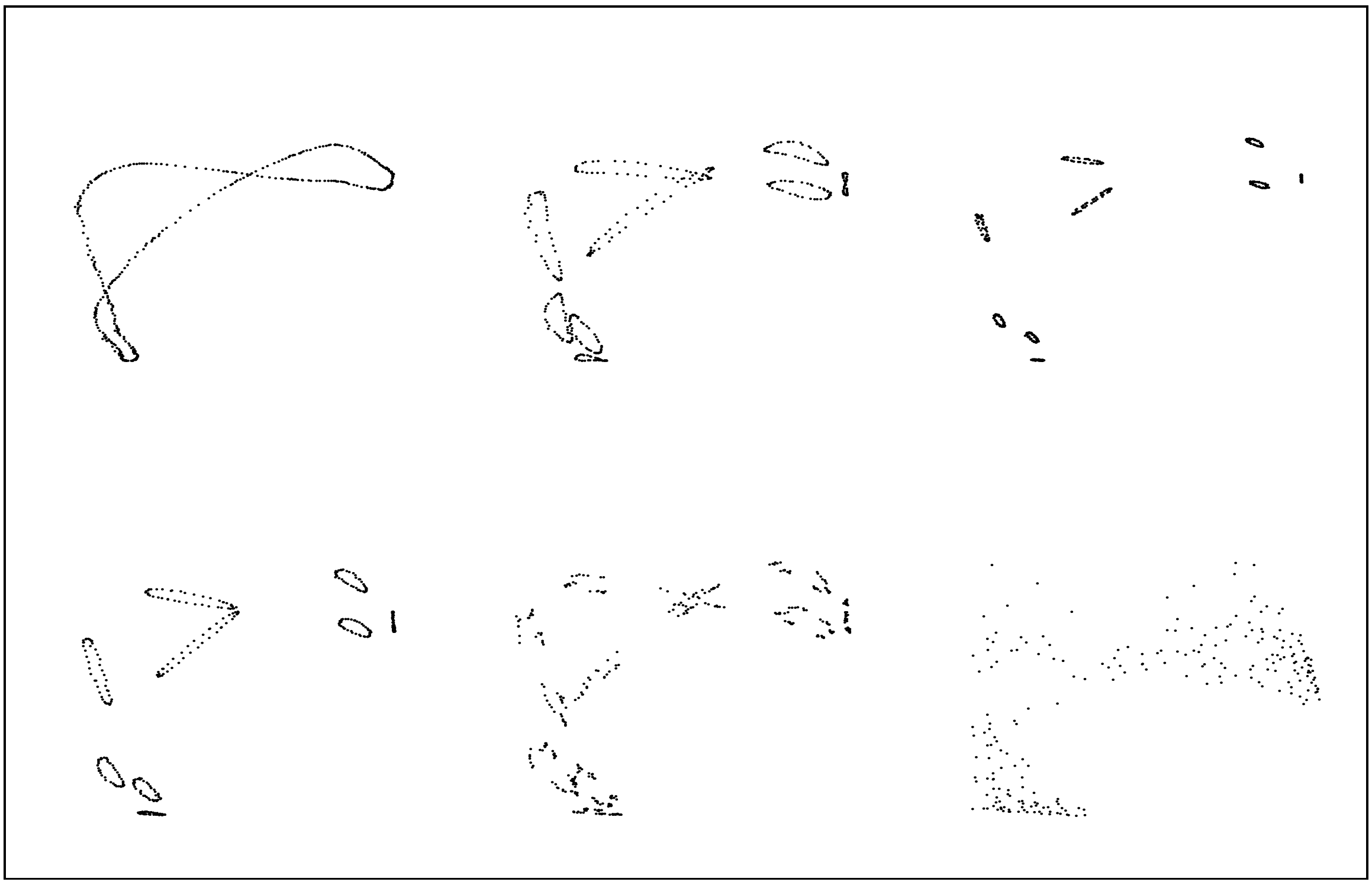

Figure 8: A route to chaos: the Poincaré surfaces of section for models with $e_0 = 0.6$ and $\Delta_0$ between 2.845 and 2.848. Read from left to right, top to bottom.

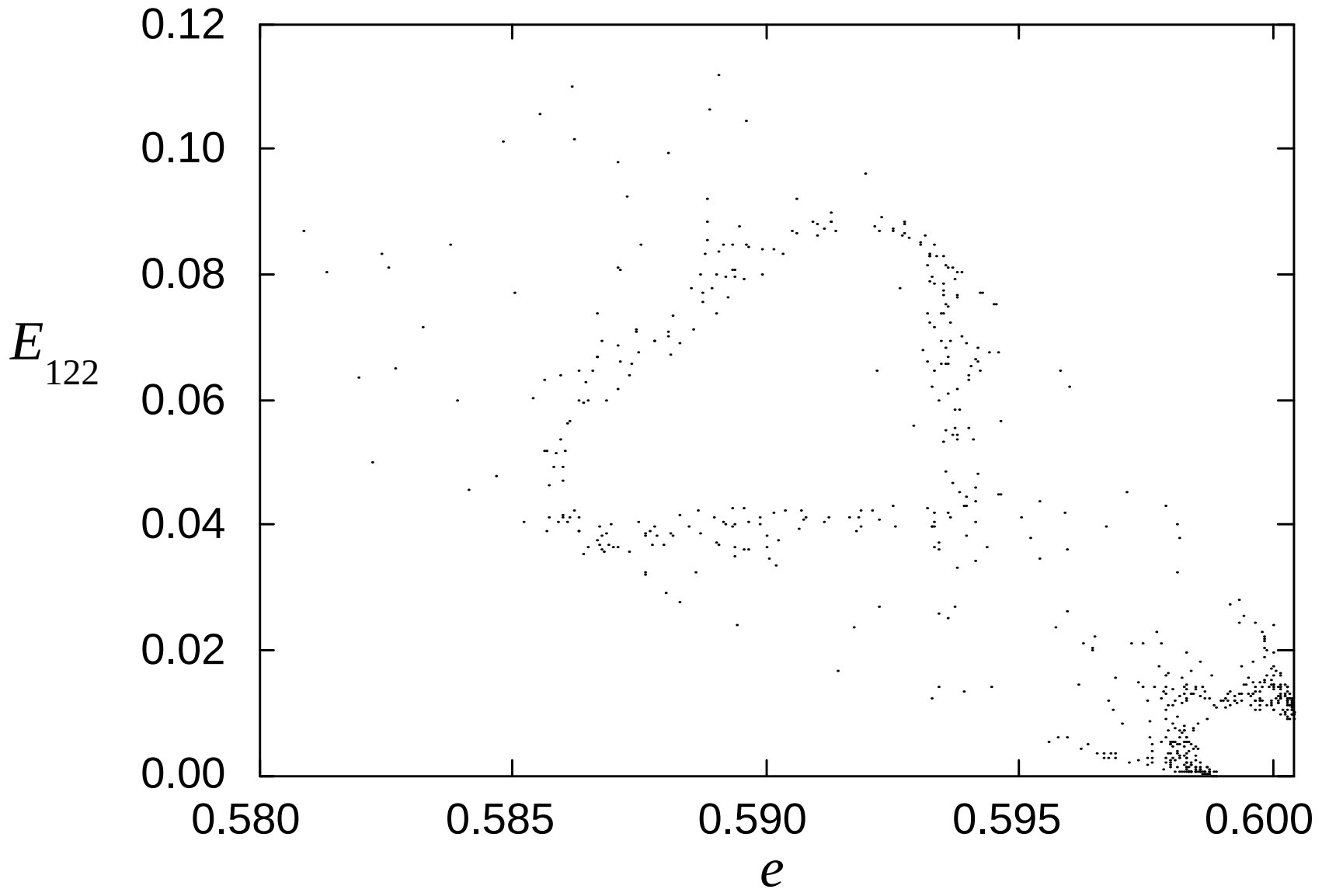

Figure 9: Chaotic and non-chaotic motion intertwined: this shows the final plot in Figure 7 (the lower right-hand corner of this figure) for longer times.

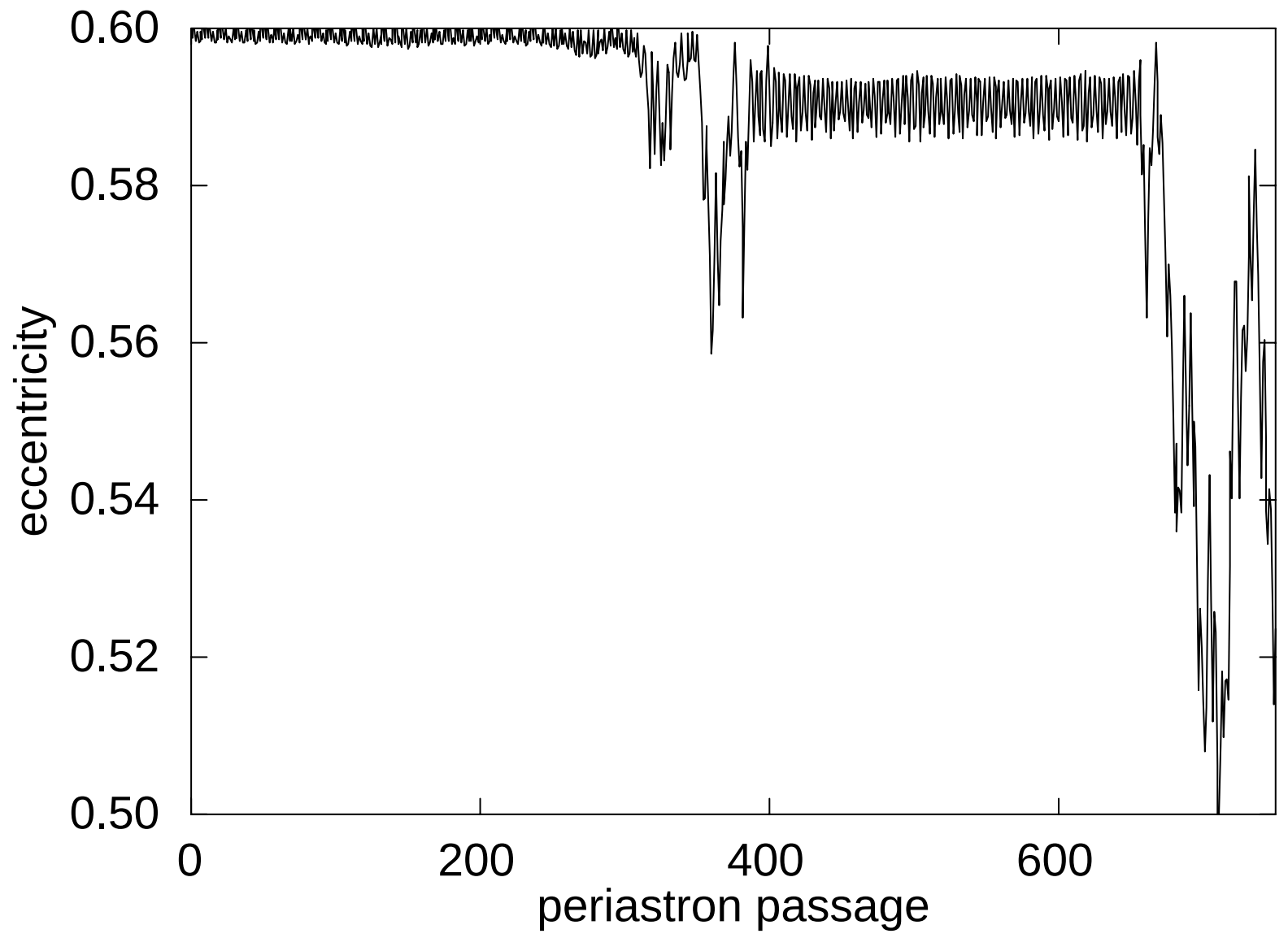

Figure 10: Chaotic and non-chaotic motion intertwined: the eccentricity versus periastron passage for the model shown in Figure 8.

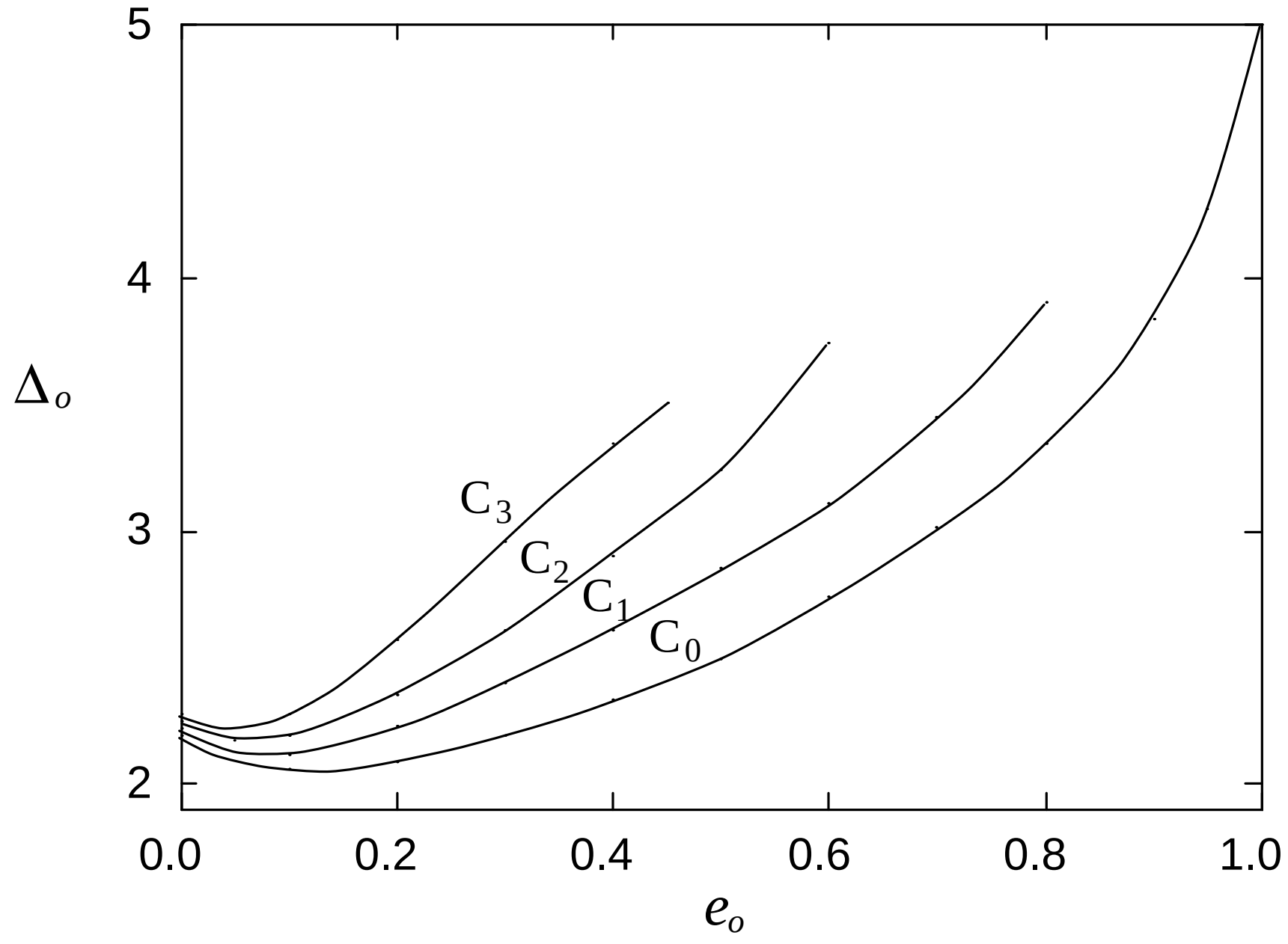

Figure 11: Chaos boundaries. A chaos boundary labeled $C_x$ is calculated by starting the polytrope off with an amount $x$ of oscillation energy. These curves must terminate at a point corresponding to zero total energy.

## 8 Discussion

We have shown that two distinctly different types of behaviour are possible when a fully self-consistent treatment of the dynamical evolution of close eccentric binaries is considered. For non-chaotic orbits, the eccentricity of the orbit is quasi-periodic, with the maximum tidal energy always remaining small. In contrast, the amount of energy transferred to the tides in a chaotic orbit can be substantial. The process should not depend on the structure of the star being polytropic, rather it is the mass distribution and hence the ability of a star to store tidal energy which is important.

One fascinating aspect of this work is that the range of periastron separations for which the tidal capture process is possible (which depends on the local velocity dispersion), *coincides almost exactly* with the range of periastron separations for which chaotic behaviour occurs. This is true for stars with mass distributions something like an $n = 1.5$ polytrope. For more centrally condensed stars, the range of periastron separations which make tidal capture possible reduces (McMillan, Taam & McDermott 1990), while the chaos boundary can be expected to drop (the tides cannot store as much energy). It remains to be seen whether these ranges coincide as they do for the present case.

In conclusion, we may say that the *tidal capture process would not be possible were it not for the existence of chaotic behaviour.*

APPENDIX

## Relationship Between the Standard and Gingold & Monaghan analyses

The standard normal mode analysis (Unno et al. 1989, p90) involves a displacement vector $\boldsymbol{\xi}$ which can be expressed as a sum of radial and poloidal components:

$$\boldsymbol{\xi} = \sum_{\mathbf{k}} \cos(\omega_{kl} t) \left( \xi_{kl}^{R}(r) \mathbf{e}_r + \xi_{kl}^{S}(r) r \boldsymbol{\nabla} \right) Y_{lm}(\theta, \varphi) \tag{20}$$

which is scaled in natural units (but see later). The velocity potential used in the present analysis may therefore be related to $\boldsymbol{\xi}$ via the (unscaled) velocity of the fluid:

$$\mathbf{v} = \frac{\alpha}{\beta} \boldsymbol{\nabla} \Phi = \left( \frac{GM_*}{R_*} \right)^{1/2} \frac{d\boldsymbol{\xi}}{dt}, \tag{21}$$

where $M_*$ and $R_*$ are the mass and radius of the star respectively. The velocity displacement may be written in terms of normal modes as (GM)

$$\Phi = -\sum_{\mathbf{k}} \sin(\omega_{kl} t) \phi_{kl}(X) Y_{lm}(\theta, \varphi) \tag{22}$$

so that

$$\boldsymbol{\nabla} \Phi = -\sum_{\mathbf{k}} \sin(\omega_{kl} t) \left\{ \frac{d\phi_{kl}}{dX} \mathbf{e}_X + \frac{\phi_{kl}}{X} X \boldsymbol{\nabla} \right\} Y_{lm}(\theta, \varphi). \tag{23}$$

Thus from equations (20), (21) and (23), we have

$$\xi_{kl}^{R} = \left( \frac{4\pi X_0}{nQ} \right)^{1/2} \frac{1}{\omega_{kl}} \frac{d\phi_{kl}}{dX} \tag{24}$$

and

$$\xi_{kl}^{S} = \left( \frac{4\pi X_0}{nQ} \right)^{1/2} \frac{1}{\omega_{kl}} \frac{\phi_{kl}}{X}. \tag{25}$$

Note that here $\boldsymbol{\xi}$ is scaled such that (Unno et al. 1989)

$$\int_0^{R_*} \rho(r) \boldsymbol{\xi}_{\mathbf{k}} \cdot \boldsymbol{\xi}_{\mathbf{k}'}^{*} \mathbf{dr} = \delta_{\mathbf{k}\mathbf{k}'} \int_0^{R_*} \rho(r) \left[ \left( \xi_{kl}^{R} \right)^2 + l(l+1) \left( \xi_{kl}^{S} \right)^2 \right] r^2 dr, \tag{26}$$

which is used by MMT, while it is more common (PT, Lee & Ostriker 1986, Ray et al. 1987) to *normalize* $\boldsymbol{\xi}$ such that

$$\int_0^{R_*} \rho(r) \boldsymbol{\xi}_{\mathbf{k}} \cdot \boldsymbol{\xi}_{\mathbf{k}'}^{*} \mathbf{dr} = \delta_{\mathbf{k}\mathbf{k}'}. \tag{27}$$

This can lead to some confusion. Note also that the factor $\rho(r)$ in the integrand of equation (27) is missing from this expression in Lee & Ostriker (1986).

Since $\xi^S$ is related by an algebraic expression to the pressure (and hence density) and gravitational potential perturbations, we may relate this to $\eta$ in a simple way (see Unno et al. 1989 and GM).

## References

Benettin, G., Galgani, L., Giorgilli, A., & Strelcyn, J. -M. 1980, Meccanica, **15**, 9

Chandrasekhar, S. 1939, Introduction to the Study of Stellar Structure (New York: Dover)

Cox, J. P. 1980, Theory of Stellar Pulsation (Princeton: Princeton University Press)

Fabian, A. C., Pringle, J. E., & Rees, M. J. 1975, MNRAS, **172**, 15P

Gingold, R. A. & Monaghan, J. J. 1980, MNRAS, **191**, 897 (GM)

Jackson, J. D. 1975, Classical Electrodynamics (New York: Wiley)

Kochanek, C. S. 1992, ApJ, **385**, 604

Lee, H. M. & Ostriker, J. P. 1986, ApJ, **310**, 176

Mardling, R. A. 1991, Chaos in Binary Stars, PhD thesis, (Monash University)

Mardling, R. A. 1994, ApJ, submitted (Paper II)

Mardling, R. A. 1994b, in preparation

Mardling, R. A. 1994c, in preparation

McMillan, S. L. W., McDermott, P. N. & Taam, R. E. 1987, ApJ, **318**, 261 (MMT)

McMillan, S. L. W., Taam, R. E. & McDermott, P. N. 1990, ApJ, **354**, 190

Press, W. H. & Teukolsky, S. A. 1977, ApJ, **213**, 183 (PT)

Rasband, S. N. 1990, Chaotic Dynamics of Non-linear Systems (New York: Wiley)

Rasio, F. A. & Shapiro, S. L. 1991, ApJ, **377**, 559

Ray, A., Kembhavi, A. K. & Antia, H. M. 1987, A&A, **184**, 164

Schwarzschild, M. 1965, Structure and Evolution of the Stars (New York: Dover)

Shimada, I. & Nagashima, T. 1979, Prog. Theor. Phys., **61**, 1605

Unno, W., Osaki, Y., & Ando, H. 1989, Nonradial Oscillations of Stars (Tokyo: University of Tokyo Press)